\documentclass[english]{article}
\usepackage[T1]{fontenc}
\usepackage[latin9]{inputenc}
\usepackage{units}
\usepackage{mathrsfs}
\usepackage{amsmath}
\usepackage{amssymb}
\usepackage{url}

\usepackage{graphics}
\usepackage{graphicx}
\usepackage{epsfig}
\usepackage{float}
\usepackage{xcolor}
\usepackage{afterpage}

\makeatletter

\providecommand{\tabularnewline}{\\}

\makeatother

\usepackage{babel}
\begin{document}

\title{Demonstrating quantum computing with the quark model}
\date{}
\author{R. M. Woloshyn\\
TRIUMF, 4004 Wesbrook Mall\\
Vancouver, British Columbia, V6T 2A3, Canada}

\maketitle
\begin{abstract}
The use of quantum computing to solve a problem in quantum mechanics
is illustrated, step by step, by calculating energies and transition
amplitudes in a nonrelativistic quark model. The quantum computations
feature the use of variational quantum imaginary time evolution implemented
using automatic differentiation to determine ground and excited states
of charmonium. The calculation of transition amplitudes is illustrated
utilizing the Hadamard test. Examples of readout and gate error mitigation
are included
\end{abstract}

\section{Introduction}

Nonrelativistic quark models with confining potentials were developed
in the 1970's \cite{PhysRevLett.34.369,PhysRevLett.34.706,PhysRevD.17.3090}. 
Doing calculations with these models is a straightforward
application of quantum mechanics. In this note the quark model is
used to illustrate the steps required to do such calculations on a
quantum computer. Often quantum computing examples are drawn from
Quantum Chemistry. However, these examples may not be completely understandable
without considerable background knowledge of chemistry. On the other
hand, the quark model is sufficiently simple that the calculations
presented here should be accessible with only a basic knowledge of
quantum mechanics.

A quantum computation of quarkonium was presented in
 \cite{https://doi.org/10.48550/arxiv.2202.03333} using
an algorithm different from the one used here. As well, only energies
of spin averaged S-wave states were considered. In this work , the
spin dependent interaction is included and a P-wave is also calculated.
The calculation of transition amplitudes is also discussed.

A feature of the calculations is the use of automatic differentiation
in implementing the variational quantum imaginary time evolution \cite{McArdle_2019}.
Automatic differentiation software constructs instructions for calculating
derivatives of functions themselves defined by software. It is implemented
in classical computing libraries such as Autograd \cite{autograd}
and Jax \cite{jax2018github} and it is
natural that this functionality should be extended to quantum computers.
The Pennylane quantum computing framework \cite{bergholm2020pennylane}
has extensive built-in automatic
differentiation capability mainly focused on machine learning applications.
However, quantum automatic differentiation does have applications
in scientific computations \cite{arrazola2021differentiable,PhysRevA.106.052429}
and will be used here as it vastly simplifies
the use of variational quantum imaginary time evolution.

The steps that will be followed to carry out the quantum computations
are enumerated here and will be discussed in subsequent sections.
\begin{enumerate}
\item Choose a quark model,\emph{ i.e}., fix the form of the potential and
the model parameters.
\item Calculate the matrix elements of the model Hamiltonian in a finite
set of basis states. Typically Harmonic oscillator states are used.
\item Rewrite the Hamiltonian matrix in terms of Pauli operators acting
on a quantum register, \emph{i.e}., on a set of qubits.
\item Determine the ground state. Variational quantum imaginary time evolution
(VQITE) implemented using automatic differentiation will be used.
\item Determine excited states.
\item Construct the operators that describe transitions between different
states, \emph{e.g}. $M1$ and $E1$ transitions between mesonic states with
different quantum numbers. The swap test and the Hadamard test will
be introduced here.
\item Check the computation on a wave function simulator and on an ideal
quantum computer simulator.
\item Check the effect of hardware noise and develop a noise mitigation
strategy.
\item Run on real quantum hardware.
\end{enumerate}
It is hoped that this note may be of use to people who are starting
to explore quantum computing for scientific applications or to instructors
teaching quantum mechanics who would like to introduce quantum computing
to their students. It is assumed that the reader has some understanding
of qubits, gates, circuits and how to setup and run quantum programs
using available quantum computing frameworks but may not yet be very
familiar with the common algorithms or applications. For the most
part, the Pennylane library \cite{bergholm2020pennylane} was used with 
Qiskit \cite{Qiskit} used for noisy simulations. No code is provided here. 
In any case, the serious reader would naturally prefer to write her own code. 

\section{Quark model}

\subsection{Schr{\"o}dinger equation}

The quark model with confining potentials 
was developed in the 1970's \cite{PhysRevLett.34.369,PhysRevLett.34.706,PhysRevD.17.3090}
after the discovery of the $J/\psi$ meson. The version adopted here
was used by Barnes \emph{et al.} \cite{PhysRevD.72.054026} in a study of 
charmonium, that is, mesons composed of a charm quark and an anti-charm quark. 
Ref. \cite{PhysRevD.72.054026} contains
an exhaustive study of many high-lying states. For illustrative purposes
in this note we restrict the calculation to the mesons $\eta_{c},$
$J/\psi$ and $h_{c}$ and their excited states. In the usual spectroscopic
notation these are the states $^{1}S_{0},$ $^{3}S_{1}$ and $^{1}P_{1}.$

The potential takes the form
\[
V(r)=-\frac{a}{r}+br+V_{s}(r)\vec{S}_{c}\cdot\vec{S}_{\bar{c}},
\]
 where $a=\tfrac{4\alpha_{s}}{3}$ and spin-dependent potential $V_{s}(r)$
is 
\[
V_{s}(r)=\frac{32\pi\alpha_{s}}{9m_{c}^{2}}\delta(r),
\]
 with $\delta(r)=\left(\nicefrac{\sigma}{\sqrt{\pi}}\right)^{3}e^{-\sigma^{2}r^{2}}$and
$m_{c}$ is the charm quark mass. Recall that the value of the spin-spin
operator $\vec{S}_{c}\cdot\vec{S}_{\bar{c}}$
is $-\tfrac{3}{4}$ in singlet states and $\tfrac{1}{4}$
in triplet states. Note also that the potential contains spin-orbit
and tensor terms. They are not relevant for the states we consider
and are omitted here. The parameters\footnote{The parameters are quoted as in Barnes \emph{et al}. in GeV units.
In the calculation it will be convenient to use the unit fm (fermi
or femtometer). Recall the conversion factor $\hbar c$ = 1 = 197.32
MeV fm.} of the model determined by Barnes \emph{et al.} \cite{PhysRevD.72.054026}
 are given in Table 1.

\begin{table}
\centering{}%
\begin{tabular}{|cccc|}
\hline 
$\alpha_{s}$ & $b$ & $m_{c}$ & $\sigma$\tabularnewline
\hline 
0.5461 & 0.1425 GeV$^{2}$ & 1.4794 GeV & 1.0946 GeV\tabularnewline
\hline 
\end{tabular}
\caption{Parameters of the quark model.}
\end{table}

For a given orbital angular momentum $l$ the radial wave function 
, written as $u(r)=R(r)/r$ in the notation of Ref. \cite{nuclear}, 
is determined by 
\begin{equation}
\label{eq:shro}
\left(-\frac{\mathrm{d^{2}}}{2\mu dr^{2}}+V(r)+\frac{l(l+1)}{2\mu r^{2}}\right)R(r)=ER(r),
\end{equation}
 where $\mu$ is the reduced mass. The meson mass is equal to the
energy eigenvalue $E$ plus 2$m_{c}.$ There are many options for
obtaining the solution of (1). The Schr{\"o}dinger equation solver 
$\mathtt{nsolve.f90}$ \cite{nsolve} was used here. 
Table 2 contains the energy eigenvalues and masses.
There is good agreement with the nonrelativistic model results given
in \cite{PhysRevD.72.054026}.

\begin{table}
\begin{centering}
\begin{tabular}{|c|c|c|c|c|c|c|}
\hline 
\multicolumn{1}{|c}{} & \multicolumn{1}{c}{$^{1}S_{0}$} &  & \multicolumn{1}{c}{$^{3}S_{1}$} &  & \multicolumn{1}{c}{$^{1}P_{1}$} & \tabularnewline
\multicolumn{1}{|c}{State} & \multicolumn{1}{c}{E {[}fm$^{-1}${]}} & Mass & \multicolumn{1}{c}{E {[}fm$^{-1}${]}} & Mass & \multicolumn{1}{c}{E {[}fm$^{-1}${]}} & Mass \tabularnewline
\hline 
1 & 0.115 & 2982 & 0.665 & 3090 & 2.826 & 3516\tabularnewline
\hline 
2 & 3.403 & 3630 & 3.613 & 3672 & 4.901 & 3934\tabularnewline
\hline 
3 & 5.496 & 4043 & 5.640 & 4072 & 6.692 & 4279\tabularnewline
\hline 
4 & 7.221 & 4384 & 7.335 & 4409 & 8.418 & 4585\tabularnewline
\hline 
\end{tabular}\caption{Quark model energy eigenvalues from Eq. (1) in units of fm$^{-1}$ and masses in units of MeV.}
\par\end{centering}
\end{table}

\subsection{Harmonic oscillator basis}

The next step in preparing the quark model calculation for a quantum
computer is to formulate the problem in matrix form. Choose a finite
set of basis states in which to represent the Hamiltonian.
The harmonic oscillator wave functions, commonly used in 
Nuclear Physics \cite{nuclear},
are a convenient choice. Like the quark model potential, the oscillator
potential is also confining and most of the calculations for matrix
elements can be done in closed form.

\begin{figure}[tb]
\centering
\includegraphics[scale=0.5,trim={1cm 1cm 2cm 1cm},clip=true,angle=0]{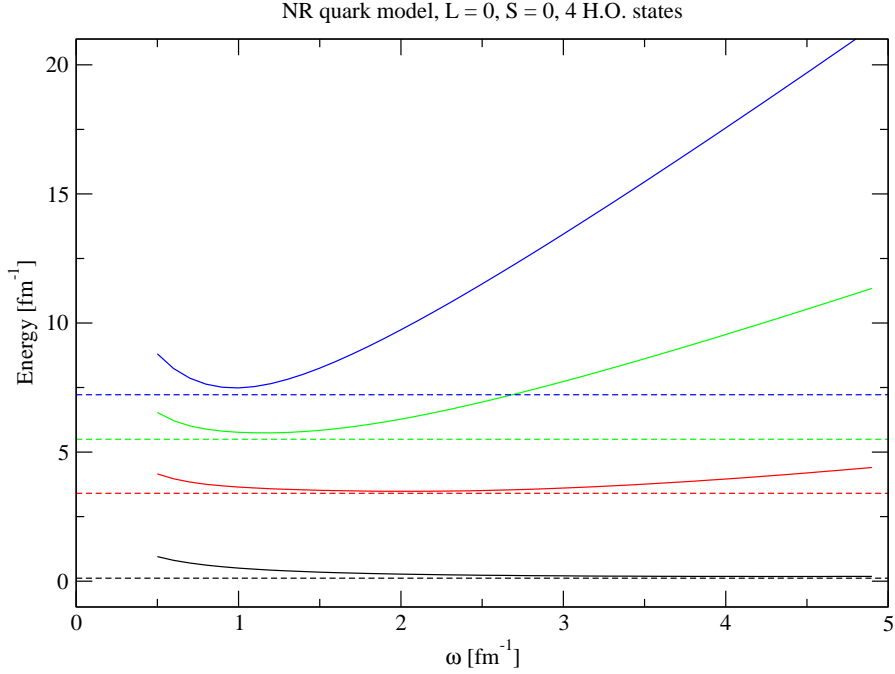}
\caption{Energy eigenvalues as function of the harmonic oscillator parameter
$\omega$ for the $^{1}S_{0}$ charmonium states from diagonalizing the 
Hamiltonian Matrix truncated to four basis states. Dashed lines are results 
from solving the Schr{\"o}dinger equation.} 
\label{omega}
\end{figure}

The Hamiltonian for the harmonic oscillator in three dimensions is
\begin{equation}
H_{HO}=-\frac{\vec{\nabla}^{2}}{2\mu}+\frac{1}{2}\mu\omega^{2}r^{2},
\end{equation}
 so the quark model Hamiltonian can be written as
\begin{equation}
\label{eq:hqm}
H_{qm}=H_{HO}+V(r)-\frac{1}{2}\mu\omega^{2}r^{2}.
\end{equation}

Recall that for quantum numbers $n,l,m$ the harmonic oscillator wave
function (unnormalized) is

\[
r^{l}e^{-\frac{1}{2}\nu r^{2}}L_{n+l-\nicefrac{1}{2}}^{l+\nicefrac{1}{2}}(\nu r^{2})Y_{lm}(\theta,\phi),
\]
 where $\nu=\mu\omega.$ The eigenenergy is $\omega(2n+l-\tfrac{1}{2})$
with $\hbar$ set equal to 1. For the present calculation the basis
is truncated to four states. The radial integrals needed to calculate
the matrix elements of (\ref{eq:hqm}) mainly involve powers of $r$ times a 
Gaussian and can be done analytically. The matrix element of the spin-dependent
term can be computed numerically.

The truncated Hamiltonian matrix depends on the oscillator parameter
$\omega.$ The energy eigenvalues calculated by diagonalizing the
Hamiltonian matrix are shown in Fig. \ref{omega} as function of $\omega$ for
the $^{1}S_{0}$ channel. The results in other channels are similar.
The dashed lines in the figure are the exact results from solving
the Schr{\"o}dinger equation. The basis truncation introduces an error
which, of course, could be reduced by enlarging the basis. For purposes
of illustrating the quantum computing process we will proceed with
only four states and choose a value of $\omega$ equal to 1.2 fm$^{-1}$.
The Hamiltonian matrix (in units of fm$^{-1}$) for the $^{1}S_{0}$
channel at this value of $\omega$ is
\begin{equation}
\label{eq:ham1s0}
\left(\begin{array}{cccc}
0.9431 & -0.8733 & -0.7690 & -0.5601\\
-0.8733 & 3.33652 & -0.5646 & -0.8648\\
-0.7690 & -0.5646 & 5.4382 & -0.1566\\
-0.5601 & -0.8648 & -0.1566 & 7.3451
\end{array}\right).
\end{equation}
 The matrices for the $^{3}S_{1}$ and $^{1}P_{1}$ channels are given
in the Appendix. The eigenvalues E of the Hamiltonian matrices at $\omega$
= 1.2 fm$^{-1}$ are in Table 3 and they will be a target of the
quantum computation.

\begin{table}
\noindent \begin{centering}
\begin{tabular}{|c|c|c|c|}
\hline 
\multicolumn{1}{|c}{} & \multicolumn{1}{c}{$^{1}S_{0}$} & \multicolumn{1}{c}{$^{3}S_{1}$} & $^{1}P_{1}$\tabularnewline
\multicolumn{1}{|c}{State} & \multicolumn{1}{c}{E {[}fm$^{-1}${]}} & \multicolumn{1}{c}{E {[}fm$^{-1}${]}} & E {[}fm$^{-1}${]}\tabularnewline
\hline 
1 & 0.395 & 0.753 & 2.783\tabularnewline
\hline 
2 & 3.506 & 3.634 & 4.875\tabularnewline
\hline 
3 & 5.664 & 5.723 & 6.765\tabularnewline
\hline 
4 & 7.546 & 7.648 & 8.767\tabularnewline
\hline 
\end{tabular}
\par\end{centering}
\caption{Energy eigenvalues from diagonalizing the truncated Hamiltonian matrices.}

\end{table}

\subsubsection{Hamiltonian in Pauli form}

Four states can be encoded in two qubits and is this is what will
be used in the quantum computation. The Hamiltonian matrix is rewritten
as the sum of Pauli terms consisting of a coefficient times a product
of Pauli operators. In the present case, two operators. In general,
the number of operators in a Pauli term equals the number of qubits. Counting 
the Identity as a Pauli operator the number of possible Pauli terms for
$n$ qubits is $4^{n}$ or 16 for the present case. Since (\ref{eq:ham1s0}) is a
real matrix only ten terms will contribute. The Hamiltonians in Pauli
form for the quark model problem are then (omitting Identity operators)
\[
c_{0}+c_{1}Z_{0}+c_{2}Z_{1}+c_{3}Z_{0}Z_{1}+c_{4}X_{0}+c_{5}X_{1}+c_{6}Z_{0}X_{1}+c_{7}X_{0}Z_{1}+c_{8}X_{0}X_{1}+c_{9}Y_{0}Y_{1},
\]
 where the subscripts on the Pauli operators $X,Y,Z$ denote the qubit
on which the operator acts. The coefficients for the $^{1}S_{0}$
Hamiltonian Eq. (\ref{eq:ham1s0}) are
\[
c=\{4.273,-2.119,-1.082,-0.129,-0.817,-0.515,-0.358,0.048,-0.562,-0.002\}.
\]
 The coefficients for $^{3}S_{1}$ are 
\[
c=\{4.439,-2.122,-1.086,-0.137,-0.655,-0.350,-0.361,0.044,-0.401,0.010\},
\]
 and for $^{1}P_{1}$
\[
c=\{5.798,-1.910,-0.964,-0.067,-0.446,0.083,-0.323,-0.064,-0.095,0.133\}.
\]

\section{Variational Quantum Imaginary Time Evolution}

There are a number of options available to calculate energy eigenvalues
by quantum computation. Quantum phase estimation 
\cite{https://doi.org/10.48550/arxiv.quant-ph/9511026,PhysRevLett.83.5162}
was developed early
on but is considered too demanding for today's limited quantum computing
resources. The variational quantum eigensolver \cite{Peruzzo_2014,PhysRevA.99.062304} 
is probably the most popular method. There are many variations of this algorithm 
\cite{PhysRevResearch.1.033062,PhysRevResearch.4.013173}
including schemes for calculating excited states. 

In this work we use an alternative, namely, imaginary time evolution.
To see why this is useful consider a Hamiltonian $H$ with eigenvalues
$\{\lambda_{i}\}$and eigenstates $\{\left|\phi_{i}\right\rangle \}.$
Then an arbitrary initial state $\left|\psi_{0}\right\rangle $ can
be expanded as $\sum_{i}\left\langle \phi_{i}\left|\right.\psi_{0}\right\rangle \left|\phi_{i}\right\rangle .$
The action of $e^{-H\tau}$ on $\left|\psi_{0}\right\rangle $ is
then 
\begin{equation}
\label{eq:ite}
e^{-H\tau}\left|\psi_{0}\right\rangle =\sum_{i}e^{-\lambda_{i}\tau}\left\langle \phi_{i}\left|\right.\psi_{0}\right\rangle \left|\phi_{i}\right\rangle .
\end{equation}
 It is clear that for sufficiently large $\tau,$ the lowest-energy
eigenstate having an overlap with $\left|\psi_{0}\right\rangle $
will dominate the sum in (\ref{eq:ite}) so that measuring the imaginary time
evolution (left hand side of (\ref{eq:ite})) in that limit will allow a 
determination of the lowest eigenvalue.

Quantum computing uses unitary operators so the non-unitary operation
(\ref{eq:ite}) can not be implemented as written. The key insight is to preserve
the norm of the state during evolution. Consider an evolution step
$\Delta\tau.$ One wants to find a unitary operator $U(\Delta\tau)$
such that 

\begin{equation}
\label{eq:qite}
\left|\psi_{\tau+\Delta\tau}\right\rangle =U(\Delta\tau)\left|\psi_{\tau}\right\rangle =\frac{e^{-H\Delta\tau}\left|\psi_{\tau}\right\rangle }{\sqrt{\left\langle \psi_{\tau}\left|e^{-2H\Delta\tau}\right|\psi_{\tau}\right\rangle }}.
\end{equation}
 At this point, one has options of how to choose the initial state
which will also affect how the unitary operator is determined. For
example, one can choose a simple to prepare state for $\left|\psi_{0}\right\rangle $,
such as a computational basis state. Then, matching a $\Delta\tau$ expansion
of different sides of (\ref{eq:qite}) gives equations for an effective unitary
operator. This is the strategy of \cite{Motta_2019,Gomes_2020}. An 
attractive feature
of this approach is that no commitment is made to the form of the
wave function.

An alternative is to use a parametrized variational ansatz for $\left|\psi_{0}\right\rangle $.
The evolution will then involve a step-by-step updating of the variational
parameters. This is the variational quantum imaginary time evolution
of \cite{McArdle_2019}
and the method that will be used here.

The variational state $\left|\psi(\vec{\theta})\right\rangle $depends
on a set of parameters $\vec{\theta}$ that are implicit functions
of $\tau$ and is produced by a chosen parametrized unitary operator
(the variational ansatz)
\[
\left|\psi(\vec{\theta})\right\rangle =U(\vec{\theta})\left|0\right\rangle .
\]
 The evolution of $\left|\psi(\vec{\theta})\right\rangle $ is governed
by the variational principle \cite{McArdle_2019,broeck}
\begin{equation}
\label{eq:vp}
\delta_{\theta}\left\Vert
\left(\partial_{\tau}+H-E_{\tau}\right)\left|\psi(\vec{\theta}(\tau))\right\rangle \right\Vert =0,
\end{equation}
 where 
\begin{equation}
\label{eq:etau}
E_{\tau}=\left\langle\psi(\vec{\theta}(\tau))\left|H\right|\psi(\vec{\theta}(\tau))\right\rangle,
\end{equation}
and $\delta_{\theta}$ denotes the variation with respect to the parameters $\vec{\theta}$.
The derivation of the equations describing the evolution of the variational
parameters is given in the Supplementary Information of Ref. \cite{McArdle_2019}.The result is 
\begin{equation}
\label{eq:theta}
\sum_{j}A_{ij}\dot{\theta}_{j}=C_{i},
\end{equation}
 with 
\begin{equation}
\label{eq:ci}
C_{i}=-\frac{\partial E_{\tau}}{\partial\theta_{i}},
\end{equation}
 and
\begin{equation}
\label{eq:aij}
A_{ij}=-\frac{\partial^{2}\left\langle \psi(\bar{\theta})\left|\right.\psi(\vec{\theta})\right\rangle }{\partial\theta_{i}\partial\theta_{j}},
\end{equation}
where $\bar{\theta}$ is used to indicate that the derivatives in
(\ref{eq:aij}) act only on the right hand side of the overlap. After solving
(\ref{eq:theta}) for $\dot{\theta}$ the parameters are updated according to $\theta(\tau+\Delta\tau)=\theta(\tau)+\Delta\tau\dot{\theta}$
where $\Delta\tau$ acts like a learning rate parameter.

\begin{figure}[tb]
\centering
\includegraphics[scale=0.45,trim={7cm 4cm 6cm 4cm},clip=true,angle=270]{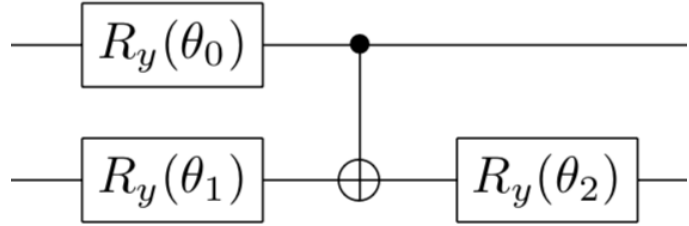}
\caption{Circuit for the variational wave function encoded in two qubits. } 
\label{ansatz}
\end{figure}

\begin{figure}[tb]
\centering
\includegraphics[scale=0.30,trim={6cm 3cm 6cm 3cm},clip=true,angle=270]{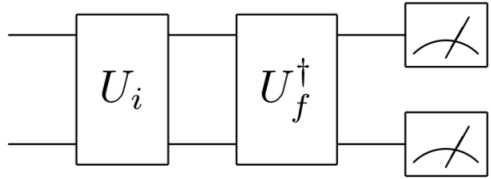}
\caption{Circuit for the wave function overlap squared $|<\psi_f|\psi_i>|^{2}$by measuring $U_f^{\dagger}U_i$.} 
\label{uud}
\end{figure}

For the computation done here with four basis states the circuit in
Fig. \ref{ansatz} is used for the variational ansatz. With three parameters this
circuit can produce any superposition of four states with real coefficients.
The circuit used for the overlap in (\ref{eq:aij}) is shown in Fig. \ref{uud}.

\nopagebreak
Although one normally associates variational calculations with ground
states, the calculation can be extended to excited states as described
in \cite{PhysRevA.99.062304}. For example, if $\left|\phi\right\rangle $ 
denotes the already determined ground state then adding
\begin{equation}
\label{eq:addE}
\alpha\left|\left\langle \phi\left|\right.\psi(\vec{\theta})\right\rangle \right|^{2}
\end{equation}
to the energy (\ref{eq:etau}) with a sufficiently large coefficient $\alpha$ 
will ensure that the variational calculation will converge to the first
excited state. This procedure can then be repeated to get even higher
lying states. The circuit in Fig. \ref{uud} is used for the overlap.

The Pennylane library is used to carry out the computations. Automatic
differentiation of quantum circuits is implemented in this framework
so calculating derivatives (\ref{eq:ci}) and (\ref{eq:aij}) is very easy. 
The Pennylane methods used for (\ref{eq:ci}) and (\ref{eq:aij}) were 
$\mathtt{pennylane.grad}$ and $\mathtt{pennylane.gradients.param\_shift\_hessian}$
respectively. Note that calculating the gradient in (\ref{eq:ci}) with the 
energy augmented by (\ref{eq:addE}) does not require any additional work on 
the programmer's part when automatic differentiation is used.

\begin{figure}[phtb]

\centerline{
\includegraphics[width=120mm,trim={0cm 2.5cm 0cm 2.5cm},clip=true]{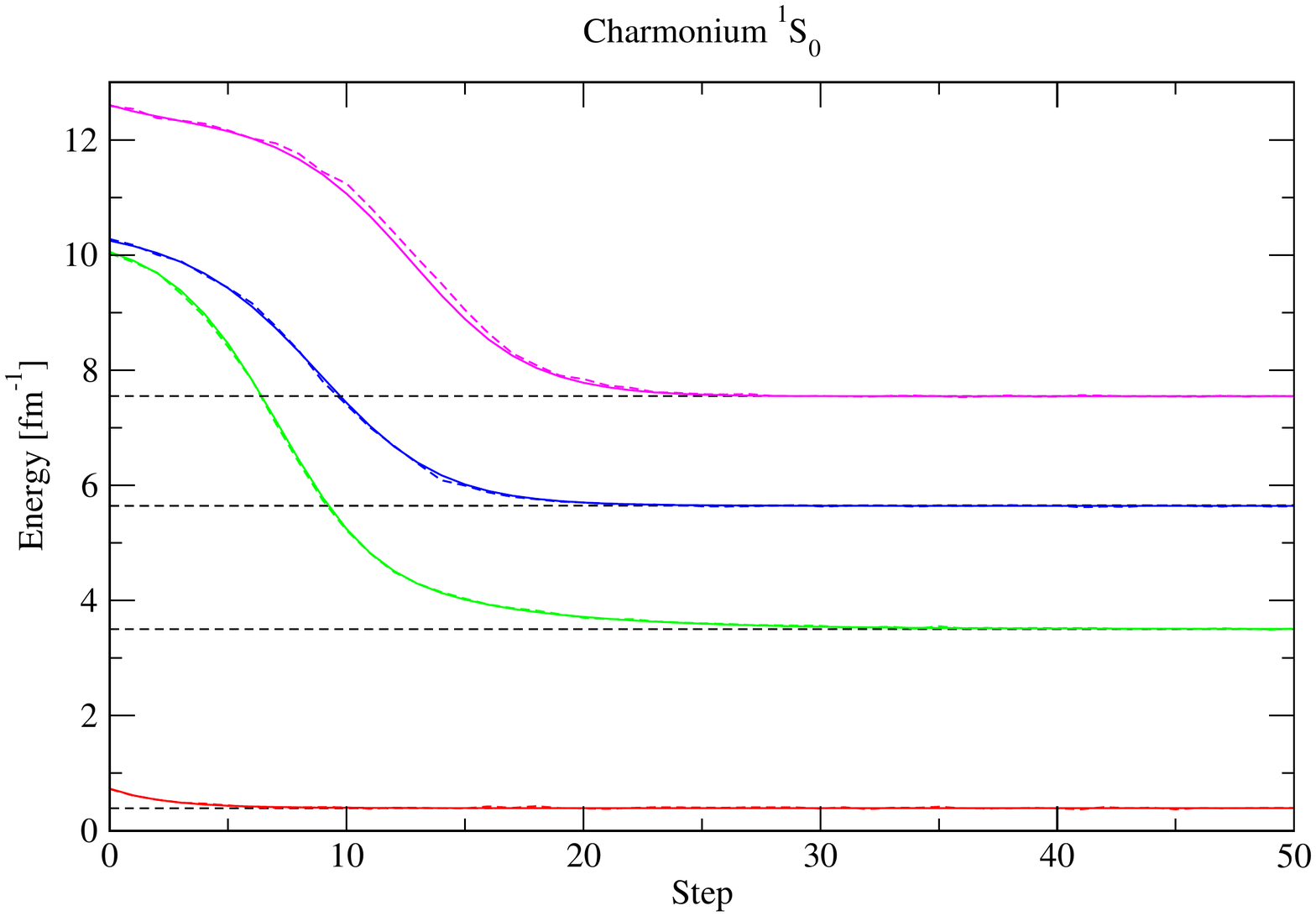}}
\caption{Evolution of the energy as a function of imaginary time
step for the $^{1}S_{0}$ channel. The solid lines are results from wave function
simulator and the coloured dashed are for an ideal quantum simulation doing
20000 shots per measurement.The horizontal dashed lines are the eigenenergies from diagonalizing
the truncated Hamiltonian matrix (\ref{eq:ham1s0}).}
\label{1S0}

\vspace*{\floatsep}


\centerline{
\includegraphics[width=120mm,trim={0cm 2.5cm 0cm 2.5cm},clip=true]{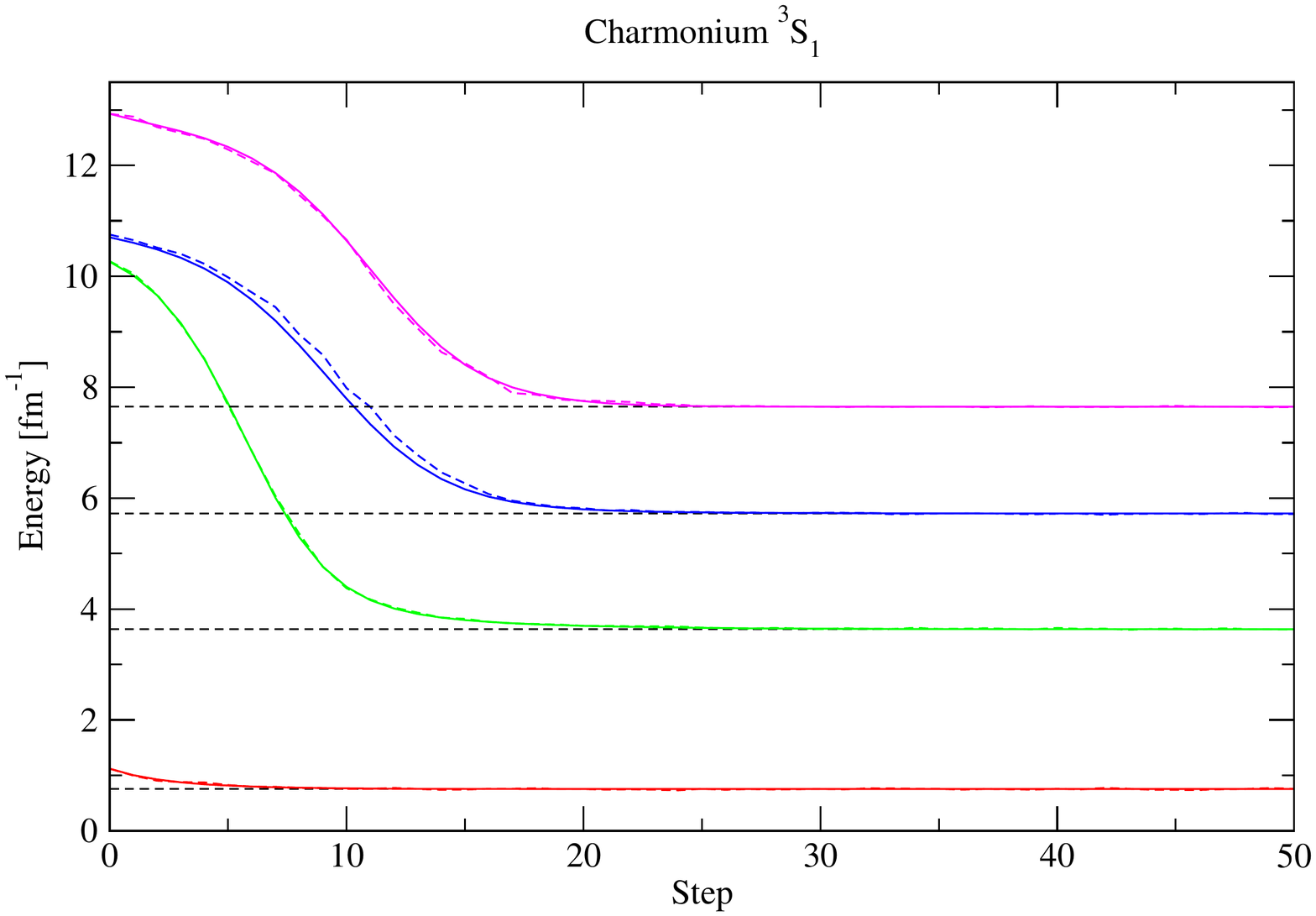}}
\caption{Evolution of the energy as a function of imaginary time
step for the $^{3}S_{1}$ channel. The solid lines are results from wave function
simulator and the coloured dashed are for an ideal quantum simulation doing
20000 shots per measurement.The horizontal dashed lines are the eigenenergies from diagonalizing
the truncated Hamiltonian matrix (\ref{eq:ham3s1}).}
\label{3S1}

\end{figure}

\begin{figure}[tb]

\centerline{
\includegraphics[width=120mm,trim={0cm 2.5cm 0cm 2.5cm},clip=true]{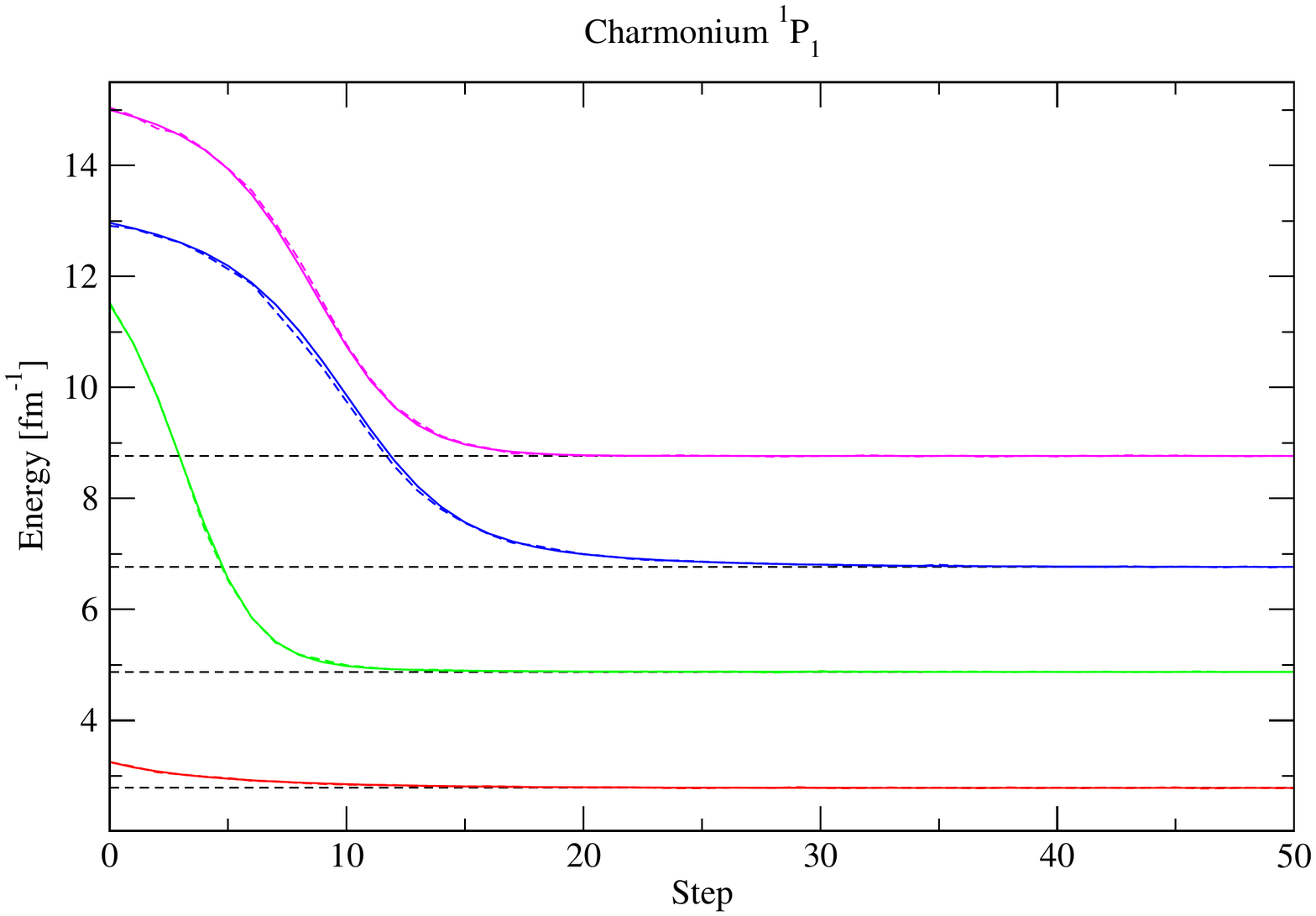}}
\caption{Evolution of the energy as a function of imaginary time
step for the $^{1}P_{1}$ channel. The solid lines are results from wave function
simulator and the coloured dashed are for an ideal quantum simulation doing
20000 shots per measurement.The horizontal dashed lines are the eigenenergies from diagonalizing
the truncated Hamiltonian matrix (\ref{eq:ham1p1}).}
\label{1P1}

\end{figure}

For the first pass at computing energies the Pennylane
$\mathtt{default.qubit}$ device
was used. With shots set equal to $\mathtt{None}$ this device acts like a 
wave function simulator, that is, it returns 
results without quantum measurement uncertainty. This provides a check on
the code. With a non-zero number of shots it simulates an ideal quantum
computer.
The step size was 0.02 and the variational parameters were
initialized to be 0.5. Figs. \ref{1S0}, \ref{3S1} and \ref{1P1} show how 
the energies in each channel change as the variational
parameters evolve. The horizontal dashed lines are the eigenenergies 
from diagonalizing the truncated Hamiltonian matrices (see Table 3).

\section{Transition amplitudes}

Having determined variational parameters that yield good ground and
excited state energies one can use the wave functions to compute other
quantities. Of particular interest for charmonium are $M1$ and $E1$
transition amplitudes. These were discussed in detail by Barnes \emph{et
al}. \cite{PhysRevD.72.054026}.

\subsection{M1 transitions}

Spin singlet and spin triplet states with the same orbital angular
momentum can transform into each other via a spin flip. In the nonrelativistic
limit the transition amplitude is just the overlap of the spatial
wave functions. Since the quark model Hamiltonian is spin dependent
there will be a finite energy transition from the higher energy triplet
ground state to the lower energy singlet ground state. Also, there
can be excited state to ground state transitions from one spin channel
to the other since the spatial wave functions are not orthogonal.
Examples of these transitions are considered here

The square of the wave function overlap can be calculated using the
$U_{i}U_{f}^{\dagger}$ circuit (Fig. \ref{uud}) that was used in the variational
calculation in Sec. 2. For comparison amplitudes were also calculated
using solutions of Eq. (\ref{eq:shro}). These are the exact results of the model
and deviations from them will show the effect of the basis truncation.

\begin{table}
\noindent \begin{centering}
\begin{tabular}{|c|c|c|c|c|}
\hline 
\multicolumn{1}{|c}{Transition} & \multicolumn{1}{c}{Exact} & \multicolumn{1}{c}{w.f. sim} & \multicolumn{1}{c}{$U_{i}U_{f}^{\dagger}$ } & Swap\tabularnewline
\hline 
$1^{3}S_{1}\rightarrow1^{1}S_{0}$ & 0.9826 & 0.9937 & 0.9939(2) & 0.9938(2)\tabularnewline
\hline 
$2^{3}S_{1}\rightarrow1^{1}S_{0}$ & 0.0107 & 0.0041 & 0.0041(1) & 0.0035(17)\tabularnewline
\hline 
$2^{3}S_{1}\rightarrow2^{1}S_{0}$ & 0.9781 & 0.9935 & 0.9936(1) & 0.9934(2)\tabularnewline
\hline 
$3^{3}S_{1}\rightarrow2^{1}S_{0}$ & 0.0061 & 0.0015 & 0.0016(1) & 0.0025(18)\tabularnewline
\hline 
$2^{1}S_{0}\rightarrow1^{3}S_{1}$ & 0.0123 & 0.0048 & 0.0047(1) & 0.0045(17)\tabularnewline
\hline 
$3^{1}S_{0}\rightarrow1^{3}S_{1}$ & 0.0025 & 0.0011 & 0.0012(1) & 0.0001(14)\tabularnewline
\hline 
\end{tabular}
\par\end{centering}
\caption{Squared amplitudes for selected $M1$ transitions. Exact results are
calculated using solutions of Eq. (1) . The last columns are results
from simulation of a ideal quantum computer using circuits in Fig. \ref{uud} 
and \ref{swap} respectively.}

\end{table}

Results for examples of squared amplitudes for $M1$ transitions are
given in Table 4. The values are either very close to one or to zero.
This reflects the orthonormality of the eigenstates. The small spin-dependent
interaction breaks the degeneracy of the spin-singlet and spin-triplet
states and shifts the overlap values slightly away from one and zero.

The column w.f. sim shows results with the Pennylane 
$\mathtt{default.qubit}$
device and shots equal to $\mathtt{None}$, essentially a wave function simulator.
The last two columns are results from simulations of an ideal quantum
computer doing 20000 shots per measurement. The results are averaged
over twenty trials. With 20000 shots, the $U_{i}U_{f}^{\dagger}$
circuit (Fig. \ref{uud}) gives excellent agreement with the wave function simulator.

\begin{figure}[tb]
\centering
\includegraphics[scale=0.50,trim={1cm 9cm 1cm 9cm},clip=true,angle=0]{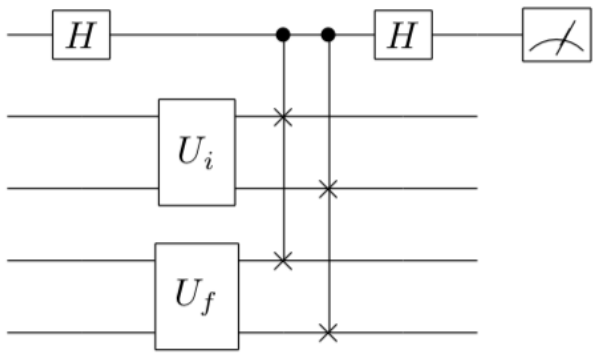}
\caption{Circuit for the wave function overlap squared $|<\psi_f|\psi_i>|^{2}$
using the swap test.} 
\label{swap}
\end{figure}

A commonly used algorithm for wave function overlap is the swap test.
The circuit is shown in Fig. \ref{swap} and the working of the swap test is
explained in the Appendix. The result of the measurement in the swap
test yields $\tfrac{1}{2}(1+\left|\left\langle \psi_{f}\left|\right.\psi_{i}\right\rangle \right|^{2})$
from which the squared overlap can be inferred. This means that when
the overlap is small, the uncertainty will be large unless the measurement
is very precise. This is reflected in the results in the last column
of Table 4. There are other variations of the swap test, for example,
given in \cite{Cincio_2018}. Investigating these is left as an exercise for the
reader.

\subsection{E1 transitions}

States with the same spin and differing by one unit in orbital angular
momentum can be connected by E1 transitions. Aside from an angular
momentum dependent factor the transition is governed by the spatial
matrix element $\left\langle u_{f}(r)\left|r\right|u_{i}(r)\right\rangle $
involving the radial wave functions of the initial and final states
\cite{PhysRevD.72.054026}. This is the quantity that will be calculated here.

Begin by calculating matrix elements of r between the radial wave
functions of the truncated harmonic oscillator basis. For P-wave to
S-wave transitions this gives (in units of fm)

\begin{equation}
\label{eq:e1ham}
\left(\begin{array}{cccc}
0.57751 & -0.4715 & 0 & 0\\
0 & 0.7455 & -0.6668 & 0\\
0 & 0 & 5.4382 & -0.9166\\
0 & 0 & 0 & 1.0002
\end{array}\right).
\end{equation}
 Written in Pauli form (\ref{eq:e1ham}) becomes
\begin{eqnarray}
\label{eq:e1ops}
 & c_{0}+c_{1}Z_{0}+c_{2}Z_{1}+c_{3}Z_{0}Z_{1}+c_{4}(X_{0}X_{1}+Y_{0}Y_{1})+c_{5}(X_{1}-iY_{1})\nonumber \\
 & +c_{6}Z_{0}(X_{1}+iY_{1})+c_{7}(X_{0}Y_{1}-Y_{0}X_{1}),
\end{eqnarray}

with coefficients
\[
c=\{0.8013,-0.1398,-0.0715,-0.0125,0.1667,-0.3220,0.0863,0.1667i\}.
\]

\begin{figure}[tb]
\centering
\includegraphics[scale=0.4,trim={7cm 0cm 7cm 0cm},clip=true,angle=270]{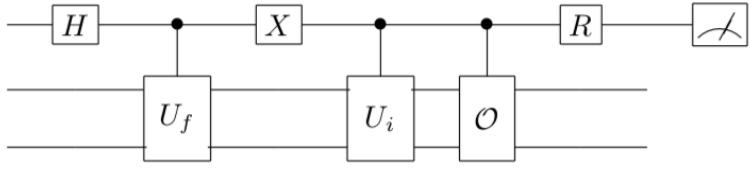}
\caption{Circuit for calculating the transition amplitude $<\psi_f|\mathcal{O}|\psi_i>$ using a Hadamard test. The gate R is R${_y}(-\pi/2)$ or 
R${_x}(\pi/2)$ for the real and imaginary parts respectively.} 
\label{hadam}
\end{figure}

The matrix elements of the Pauli terms in (\ref{eq:e1ops}) between different 
states are calculated using a Hadamard test circuit. It is shown in Fig.
\ref{hadam} and its operation is described in the Appendix. The rotation operator
R is chosen such that the measurement yields either the expectation
value of $\sigma_{x}$ or $\sigma_{y}$ for the first qubit from which
the real or imaginary part of the operator $\mathcal{O}$ matrix element
value can be inferred.

\begin{table}
\noindent \begin{centering}
\begin{tabular}{|c|c|c|c|c|}
\hline 
\multicolumn{1}{|c}{Transition} & \multicolumn{1}{c}{Exact} & \multicolumn{1}{c}{w.f. sim} & \multicolumn{1}{c}{1000} & 10000\tabularnewline
\hline 
$1^{1}P_{1}\rightarrow1^{1}S_{0}$ & 0.3490 & 0.3925 & 0.3929(37) & 0.3900(13)\tabularnewline
\hline 
$2^{1}P_{1}\rightarrow2^{1}S_{0}$ & 0.5900 & 0.7174 & 0.7229(42) & 0.7172(12)\tabularnewline
\hline 
$2^{1}S_{0}\rightarrow1^{1}P_{1}$ & 0.5757 & 0.5406 & 0.5433(73) & 0.5409(24)\tabularnewline
\hline 
$3^{1}S_{0}\rightarrow2^{1}P_{1}$ & 0.8873 & 0.8227 & 0.8174(49) & 0.8201(18)\tabularnewline
\hline 
$3^{1}S_{0}\rightarrow1^{1}P_{1}$ & 0.0311 & 0.0448 & 0.0449(65) & 0.0448(23)\tabularnewline
\hline 
\end{tabular}
\par\end{centering}
\caption{Examples of E1 transition amplitudes. The results in the last two
columns were calculated using a Hadamard test circuit with 1000 and
10000 shots per measurement respectively. }
\end{table}

The results for the radial matrix elements (in units of fm) for some
examples of $E1$ transitions are shown in Table 5. Just as for the $M1$
 overlaps, the results calculated with the exact wave functions, i.e., solutions
of (\ref{eq:shro}), are also tabulated. The truncated basis calculation reproduces
the exact results reasonably well and with 10000 shots per measurement
the quantum simulation agrees very well with the wave function simulator
results. 

\section{Noisy simulation}

So far only ideal quantum simulations have been considered. The uncertainties
that arise are due to the stochastic nature of quantum measurement
and they can be reduced by increasing the number of times measurements
are carried out. On a real quantum device there will also systematic
errors due to noise in the hardware. There can be errors, referred
to a gate errors, which affect the unitary operations that the circuit
carries out. In addition, qubits may be mis-measured,\emph{ i.e}.,
readout error. 

To get an idea of how a computation may be affected hardware noise
one can introduce a noise model in the quantum simulator. The Qiskit framework
is very useful for doing this since it provides noise models based
on actual IBM quantum hardware as well a capability of creating custom
noise models. It also provides some tools for mitigating or correcting
hardware errors. Pennylane, which is the framework used in this work,
has cross-platform capability through the use of plug-ins, so that
we can include Qiskit resources within our code. 

\begin{figure}[tb]
\centerline{
\includegraphics[width=120mm,trim={0cm 3cm 0cm 3cm},clip=true]{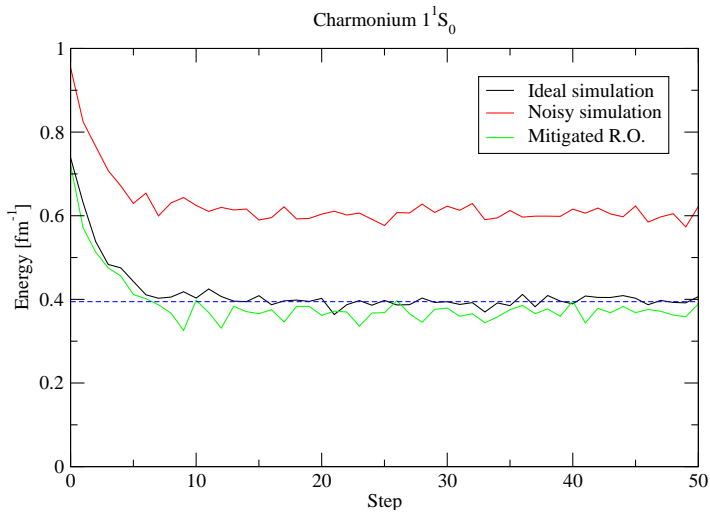}}
\caption{Evolution of the energy as a function of imaginary time
step for the $^{1}S_{0}$ ground state. Results of ideal, noisy and readout error
mitigated simulations are shown.The horizontal dashed line is the eigenenergy 
from diagonalizing
the truncated Hamiltonian matrix (\ref{eq:ham1p1}).}
\label{1S}

\end{figure}

\begin{figure}[phtb]

\centerline{
\includegraphics[width=120mm,trim={0cm 2.5cm 0cm 2.5cm},clip=true]{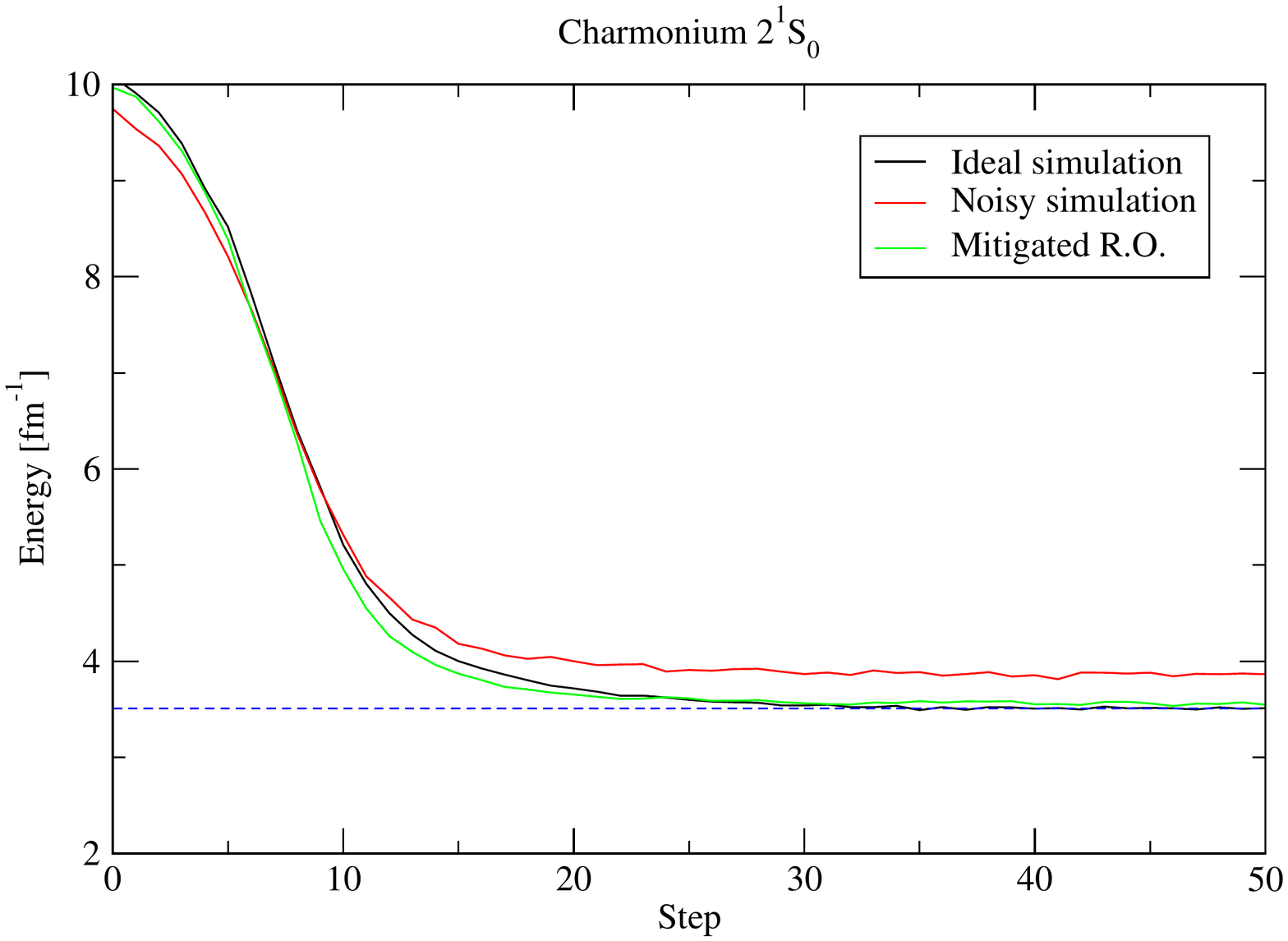}}
\caption{Evolution of the energy as a function of imaginary time
step for the $^{1}S_{0}$ first excited state. Results of ideal, noisy and readout error
mitigated simulations are shown.The horizontal dashed line is the eigenenergy 
from diagonalizing
the truncated Hamiltonian matrix (\ref{eq:ham1s0}).}
\label{2S}

\vspace*{\floatsep}

\centerline{
\includegraphics[width=120mm,trim={0cm 2.5cm 0cm 2.5cm},clip=true]{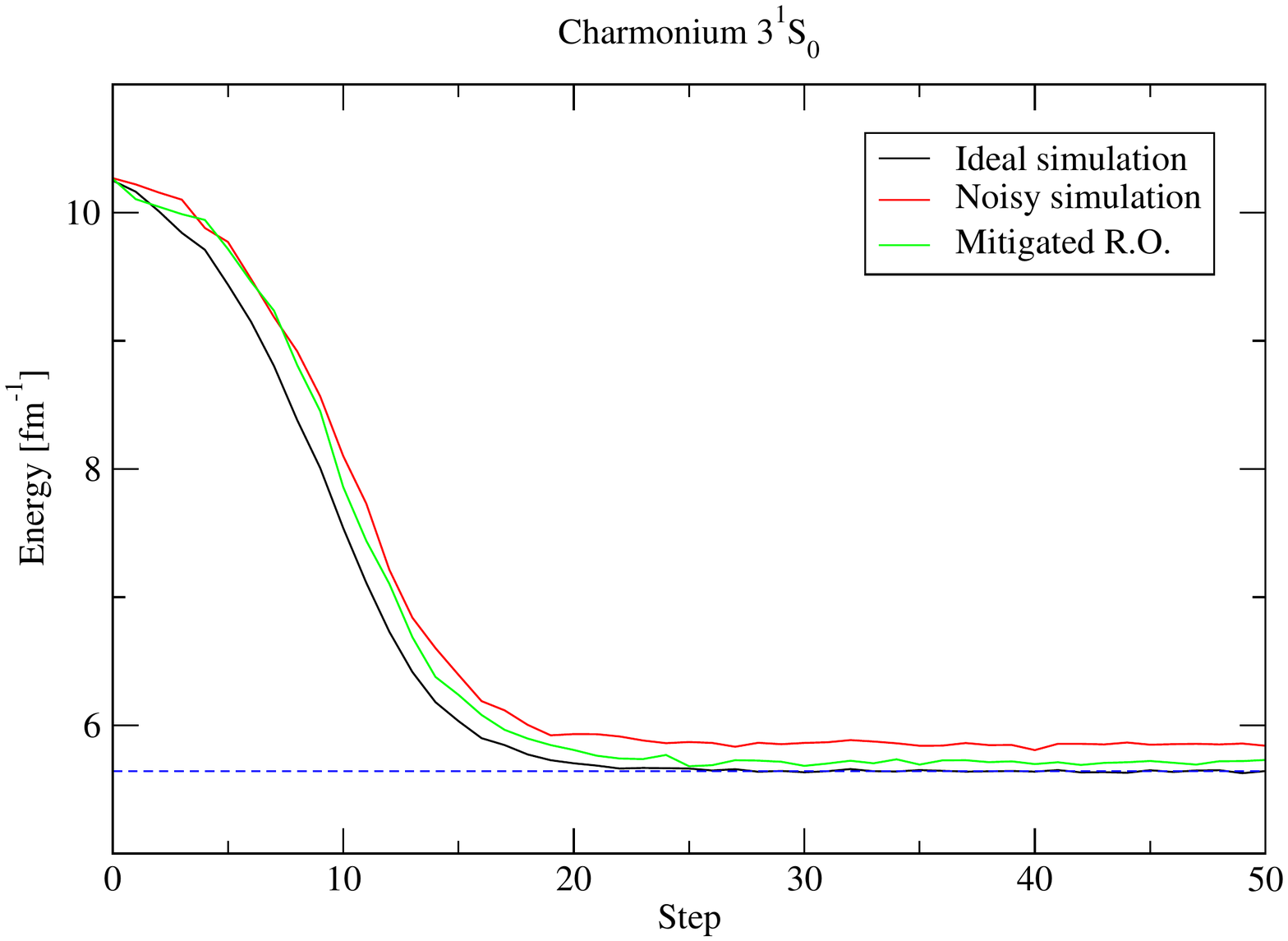}}
\caption{Evolution of the energy as a function of imaginary time
step for the $^{1}S_{0}$ second excited state. Results of ideal, noisy and readout error
mitigated simulations are shown.The horizontal dashed line is the eigenenergy 
from diagonalizing
the truncated Hamiltonian matrix (\ref{eq:ham1s0}).}
\label{3S}

\end{figure}

As a first example the mitigation of readout errors in the calculations
of energies is considered. Results for the $^{1}S_{0}$ channel are
shown here, Figs \ref{1S}, \ref{2S} and \ref{3S}. 
Results for the other channels should be similar. The
noise model used is for the device $\mathtt{ibmq\_manila}$. After downloading
the noise model\footnote{An IBM Quantum account is needed to download the noise model. An alternative
which does not need an account is to use a fake device, \emph{eg.},
FakeManila, which is included in the Qiskit software.} a calibration run is 
done to determine the readout fidelity of the
qubits that will be measured in the actual calculations. When the
calculations are done the calibration matrix is applied to the measurement
before the computing final results. The procedure used here is described
in \cite{romit}. Note that for illustration purposes, only readout error
was included in the noise model. Mitigating gate errors when using
automatic differentiation \cite{PhysRevA.106.052429} goes beyond the scope 
of this pedagogical note. 

The next example is the effect of readout error on the wave function
squared overlaps that govern $M1$ transitions. Results, calculated
using the circuit in Fig. \ref{uud}, are in Table 6.

\begin{table}
\noindent \begin{centering}
\begin{tabular}{|c|c|c|c|}
\hline 
\multicolumn{1}{|c}{Transition} & \multicolumn{1}{c}{No noise} & \multicolumn{1}{c}{With R.O. error} & Mitigated\tabularnewline
\hline 
$1^{3}S_{1}\rightarrow1^{1}S_{0}$ & 0.9939(2) & 0.9385(1) & 0.9813(1)\tabularnewline
\hline 
$2^{3}S_{1}\rightarrow1^{1}S_{0}$ & 0.0041(1) & 0.0596(1) & 0.0320(1)\tabularnewline
\hline 
$2^{3}S_{1}\rightarrow2^{1}S_{0}$ & 0.9936(1) & 0.9386(1) & 0.9808(1)\tabularnewline
\hline 
$3^{3}S_{1}\rightarrow2^{1}S_{0}$ & 0.0016(1) & 0.0270(1) & 0.0036(1)\tabularnewline
\hline 
$2^{1}S_{0}\rightarrow1^{3}S_{1}$ & 0.0047(1) & 0.0614(1) & 0.0327(1)\tabularnewline
\hline 
$3^{1}S_{0}\rightarrow1^{3}S_{1}$ & 0.0012(1) & 0.0526(1) & 0.00008(4)\tabularnewline
\hline 
\end{tabular} 
\par\end{centering}
\caption{Calculations of squared amplitudes for selected $M1$ transitions
showing the effect of readout error and mitigation.}

\end{table}

The effects of readout error are substantial. For the large overlaps,
mitigating readout errors brings the results close to the ideal noise-free
results. When the overlap is small, mitigating the read
out error provides
an improvement but not enough to provide a reliable estimate of the
true noise-free value. 

\begin{figure}[tb]
\centerline{
\includegraphics[width=120mm,trim={0cm 3cm 0cm 3cm},clip=true]{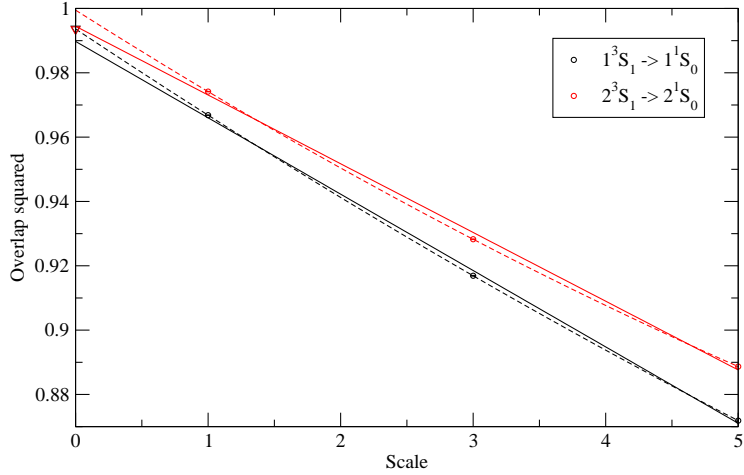}}
\caption{Wave function overlap squared for selected $M1$ transitions having a
large amplitude as function of circuit unitary folding scale. The solid and
dashed lines are linear and second-order polynomial zero-noise extrapolations.
The results of noise-free simulations are shown by triangles at scale equal 0.}
\label{zneL}

\end{figure}

\begin{figure}[phtb]

\centerline{
\includegraphics[width=120mm,trim={0cm 3cm 0cm 3cm},clip=true]{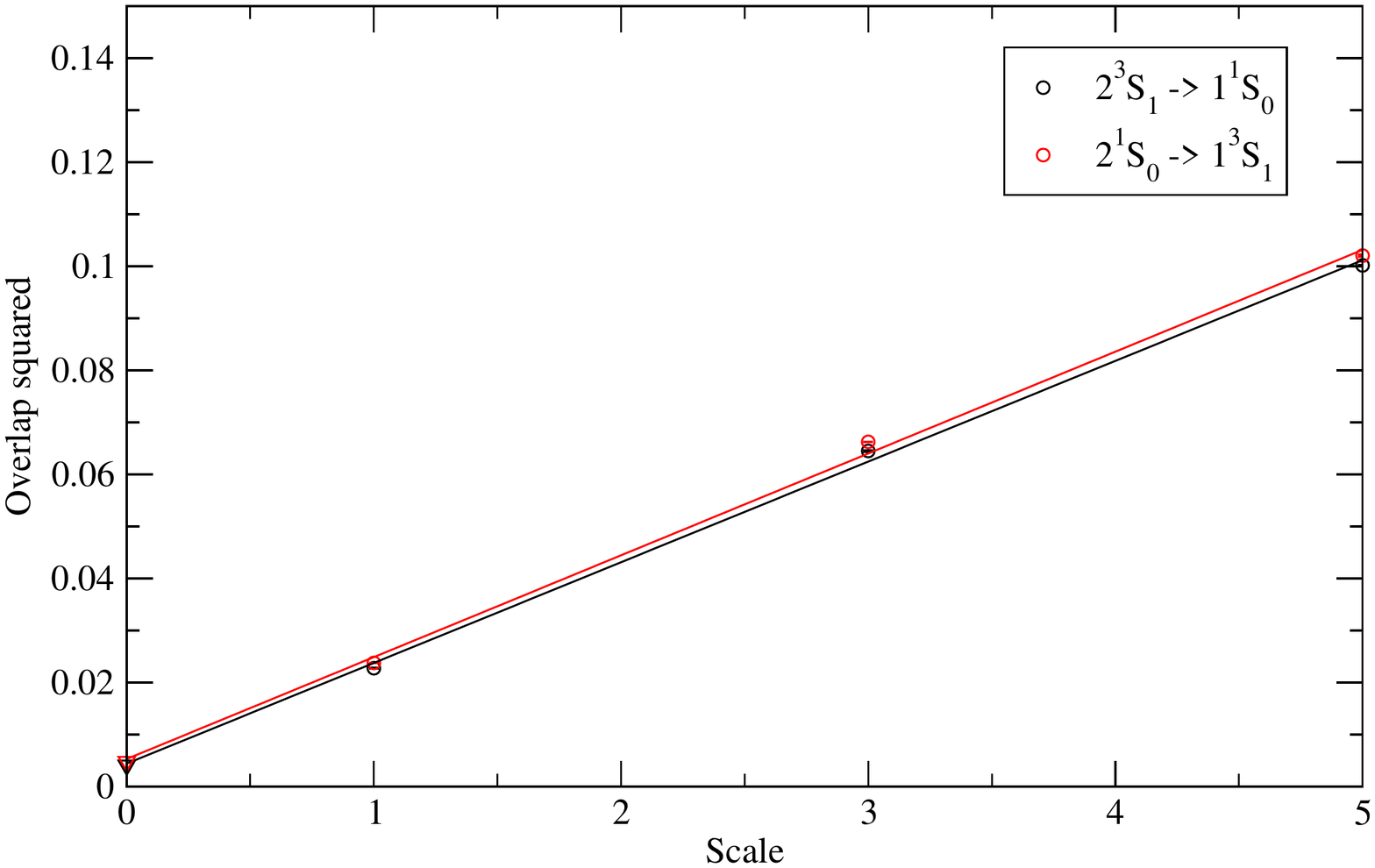}}
\caption{Wave function overlap squared for selected $M1$ transitions having a
small amplitude as function of circuit unitary folding scale. The solid lines 
are linear zero-noise extrapolations.
The results of noise-free simulations are shown by triangles at scale equal 0.}
\label{zneM}


\vspace*{\floatsep}

\centerline{
\includegraphics[width=120mm,trim={0cm 3cm 0cm 3cm},clip=true]{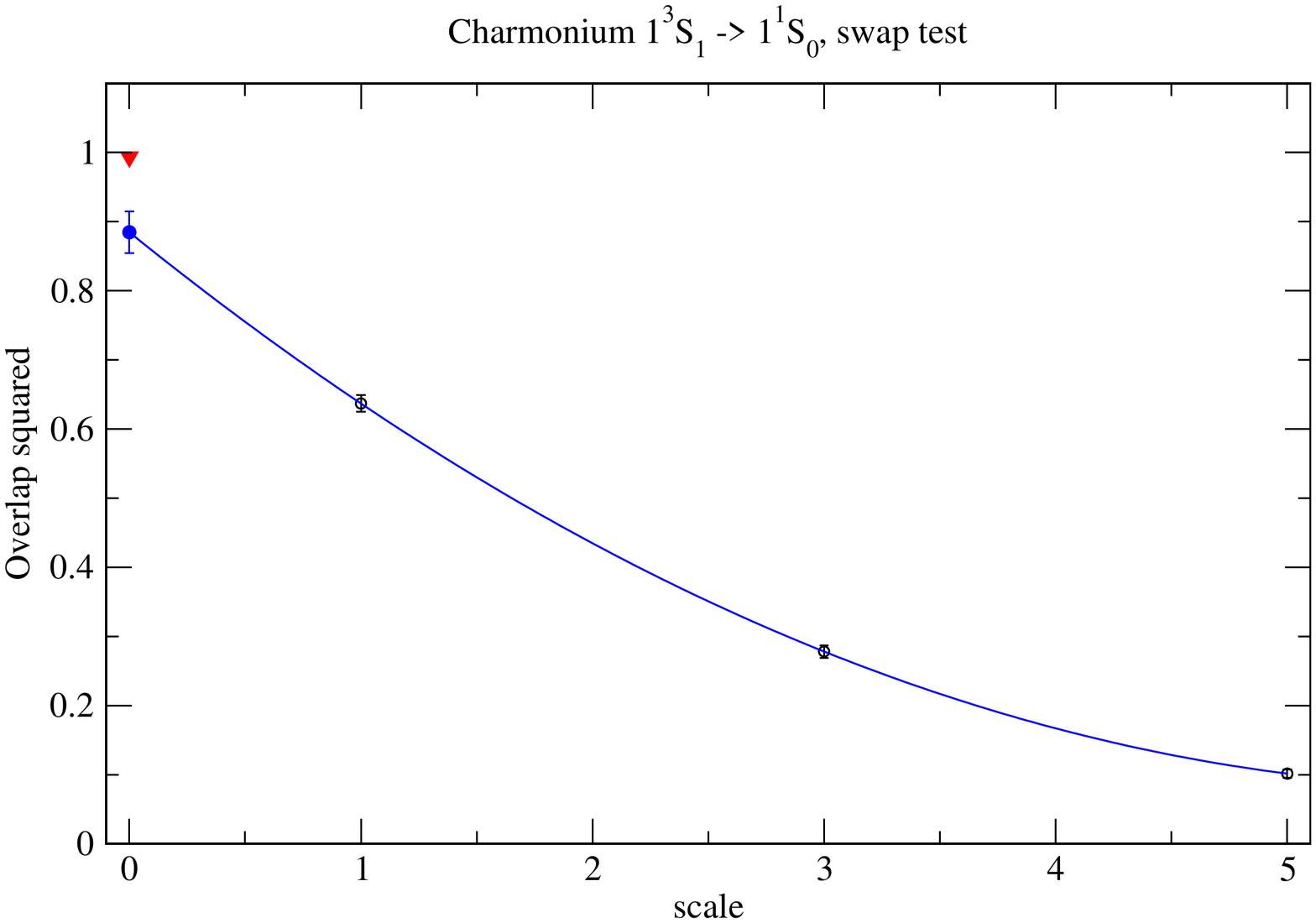}}
\caption{Wave function overlap squared describing the spin-triplet to
spin-singlet $M1$ transition as a function of unitary folding scale. The line is
a second-order polynomial extrapolation.
The result of a noise-free simulation is shown by the triangle at scale equal 0.}
\label{zneSwap}

\end{figure}

There are many suggestions of how to mitigate gate errors and it is
still an active research area
\cite{PhysRevX.7.021050,krebsbach2022optimization,PhysRevD.106.074502,
LaRose_2022,larose2022error,https://doi.org/10.48550/arxiv.2212.03937}
One of the first proposals was the
use of zero-noise extrapolation. The general idea is to calculate
results with gate noise artificially increased and then extrapolate
the results down to zero noise. One way to implement this scheme is
so-called unitary folding. Let $\mathscr{U}$ denote the circuit to
which gate error mitigation is to be applied. Then the circuits $\mathscr{UU^{\dagger}U}$,
$\mathscr{UU^{\dagger}UU^{\dagger}U},$ etc. will have increased gate
errors although they would give identical results on an ideal noise-free
quantum device. Results with different levels of $\mathscr{U^{\dagger}U}$
insertions can be extrapolated to estimate a noise-free value. This
scheme was implemented for selected $M1$ transitions calculated using
the circuit Fig. \ref{uud}. The results are shown in the 
Fig. \ref{zneL} and \ref{zneM}. Both linear and
second-order polynomial extrapolations were used. However, for the
transitions in Fig. \ref{zneM} second-order polynomial extrapolation led to
unphysical, \emph{i.e}., negative values, at zero noise so that extrapolation
is omitted. Scale equal 1 means calculation with no unitary folding,
scale equal 3 means an insertion of the circuit and its adjoint and
so on.

The final example will be gate error mitigation for the swap test
circuit Fig. \ref{swap}. Generally multi-qubit gates are more susceptible to 
errors than single quibit gates so one would expect that the swap test circuit
will have much bigger errors than the circuit in Fig. \ref{uud}. On a noisy
device it will only be feasible to calculate the transitions with
amplitude near one. An example of mitigation of gate errors by zero
noise extrapolation in a calculation using the swap test circuit is
shown in Fig. \ref{zneSwap}. The effect of gate errors is very large compared
to Fig. \ref{zneL} where the overlap circuit Fig. \ref{uud} was used. 
Nonetheless, an extrapolation can be made which gets considerably closer to , 
but not quite in agreement with, the noise-free value. The uncertainty
in the extrapolated value was obtained using a bootstrap procedure.

Error mitigation of $E1$ transition amplitudes is left for the reader.

\section{Quantum hardware}

For completeness, an example of a computations done on a real quantum
device will be given in this section. The Pennylane software library was 
used in this work so using the Pennylane-qiskit plugin allows
code to be run on IBM Quantum hardware by setting the Pennylane device 
to be $\mathtt{qiskit.ibmq}$ with backend specified as the IBM Quantum
device. 

In the previous section use was made of the ibmq\_manila noise model
running on a quantum simulator. The first example will be to compare
the real ibmq\_manila device with the simulated one. Since the author's access
to quantum resources is limited, the comparison will be made for the
simplest calculation.\emph{ i.e}., wave function overlap squared using 
circuit Fig. \ref{uud}.

\begin{table}
\begin{centering}
\begin{tabular}{|c|c|c|c|}
\hline 
\multicolumn{1}{|c}{} & \multicolumn{1}{c}{Noise-free } & \multicolumn{1}{c}{Simulated} & \multicolumn{1}{c|}{Real}\tabularnewline
\multicolumn{1}{|c}{Transition} & \multicolumn{1}{c}{simulation} & \multicolumn{1}{c}{ibmq\_manila} & \multicolumn{1}{c|}{ibmq\_manila}\tabularnewline
\hline 
$1^{3}S_{1}\rightarrow1^{1}S_{0}$ & 0.9939(2) & 0.9414(2) & 0.9314(97)\tabularnewline
\hline 
$2^{3}S_{1}\rightarrow1^{1}S_{0}$ & 0.0041(1) & 0.0432(1) & 0.0522(40)\tabularnewline
\hline 
$2^{3}S_{1}\rightarrow2^{1}S_{0}$ & 0.9936(1) & 0.9470(1) & 0.9253(118)\tabularnewline
\hline 
$3^{3}S_{1}\rightarrow2^{1}S_{0}$ & 0.0016(1) & 0.0223(1) & 0.0253(17)\tabularnewline
\hline 
$2^{1}S_{0}\rightarrow1^{3}S_{1}$ & 0.0047(1) & 0.0459(1) & 0.0509(37)\tabularnewline
\hline 
$3^{1}S_{0}\rightarrow1^{3}S_{1}$ & 0.0012(1) & 0.0542(1) & 0.0640(48)\tabularnewline
\hline 
\end{tabular}\caption{Squared amplitudes for selected $M1$ transitions comparing
results obtained on the real ibmq\_manila device with noise-free and noisy
simulations.}
\par\end{centering}
\end{table}

\begin{figure}[tb]
\centerline{
\includegraphics[width=120mm,trim={0cm 3cm 0cm 3cm},clip=true]{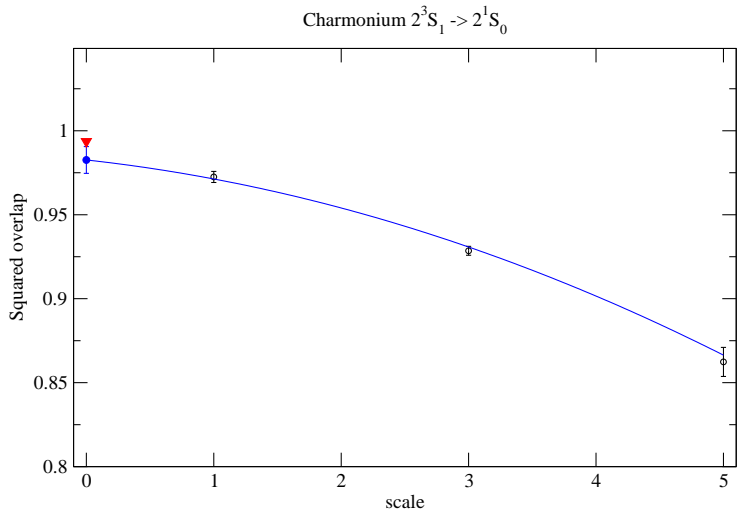}}
\caption{Wave function overlap squared describing the $2^{3}S_{1}\rightarrow2^{1}S_{0}$
transition as a function of unitary folding scale. The line is
a second-order polynomial extrapolation. The result of a noise-free simulation 
is shown by the triangle at scale equal 0.}
\label{zneMit}

\end{figure}

Squared amplitudes for selected $M1$ transitions calculated without error mitigation 
on the ibmq\_manila device are shown in the last column of Table 7. Results are averaged
over ten trials doing 20000 shots per measurement. For comparison
the results from an ideal noise-free simulation and from simulated results using the 
ibmq\_manila noise model are also included. The noisy simulation and real hardware results
are similar. The main difference being the larger uncertainty in the hardware results
reflecting considerable fluctuation in results of different trails carried out over a 
period of a few days.

The final example is a calculation of the wave function overlap squared describing 
the $2^{3}S_{1}\rightarrow2^{1}S_{0}$ transition doing both readout and gate error
mitigation. The computed results using different unitary foldings are indicated by
open circles in Fig. \ref{zneMit}. The solid circle is the zero-noise extrapolated value
with uncertainty obtained by a bootstrap analysis.

\section{Summary}

The calculation of energies and transition amplitudes in a nonrelativistic
quark model is used to demonstrate the steps required to carry out
quantum computations. Ground and excited states energies of charmonium
were calculated using variational quantum imaginary time evolution.
The algorithm was implemented using automatic differentiation within
the Pennylane quantum computing framework. Amplitudes for selected
$M1$ and $E1$ transitions were calculated illustrating the use of
the swap test and the Hadamard test.

Effects of readout and gate errors were investigated through the use
of simulations with noise models accessed using the Qiskit framework.
Readout error correction and zero-noise extrapolation were shown to
improve results but in cases, for example, where the the transition
amplitude was very small error mitigation was not sufficient to yield
reliable results.

Some examples of computations carried out on the IBM ibmq\_manila 
device were presented. It should be noted that with Pennylane's 
cross-platform capability these calculations can be ported easily 
to other kinds of quantum computing hardware. 

\section*{Acknowledgement}

Thanks to Olivia Di Matteo and Billy Jones for their helpful comments.
We acknowledge the use of the IBM Quantum
services for this work. The views expressed
are those of the author, and do not reflect the official
policy or position of IBM or the IBM Quantum team.
TRIUMF receives federal funding via a contribution agreement with
the National Research Council of Canada.

\section*{Appendix}

\subsection*{A.1 Oscillator basis Hamiltonians}

The truncated Hamiltonian for $^{3}S_{1}$ at $\omega$= 1.2 fm$^{-1}$

\begin{equation}
\label{eq:ham3s1}
\left(\begin{array}{cccc}
1.0946 & -0.7114 & -0.6111 & -0.4112\\
-0.7114 & 3.5406 & -0.3910 & -0.6989\\
-0.6111 & -0.3910 & 5.6122 & -0.0119\\
-0.4112 & -0.6989 & -0.0119 & 7.5104
\end{array}\right),
\end{equation}
 and for $^{1}P_{1}$

\begin{equation}
\label{eq:ham1p1}
\left(\begin{array}{cccc}
2.8561 & -0.2395 & -0.3827 & -0.2282\\
-0.2395 & 4.919 & -0.0373 & -0.5097\\
-0.3827 & -0.0373 & 6.8114 & 0.4058\\
-0.2282 & -0.5097 & 0.4058 & 7.5104
\end{array}\right).
\end{equation}
 The matrix elements are in units of fm$^{-1}.$

\subsection*{A.2 Overlap circuits}

The swap test and Hadamard test circuits for calculating the transition
amplitudes are shown in Figs. \ref{swap} and \ref{hadam}. A brief explanation 
of how these circuits work is given here. 

First, consider the swap test. The quantum registers for the ancillary
qubit and the inital and final wave functions are initially in the
state $\left|0\right\rangle _{0}\left|0\right\rangle _{1}\left|0\right\rangle _{2}$
and then operations as listed below are carried out.
\begin{center}
\begin{tabular}{cc}
Operation & State\tabularnewline
\hline 
 & $\left|0\right\rangle _{0}\left|0\right\rangle _{1}\left|0\right\rangle _{2}$\tabularnewline
Hadamard(0) & \tabularnewline
 & $\frac{1}{\sqrt{2}}\left(\left|0\right\rangle _{0}\left|0\right\rangle _{1}\left|0\right\rangle _{2} +\left|1\right\rangle _{0}\left|0\right\rangle _{1}\left|0\right\rangle _{2}\right)$\tabularnewline
$U_{i}(1)$ & \tabularnewline
 & $\frac{1}{\sqrt{2}}\left(\left|0\right\rangle _{0}\left|i\right\rangle _{1}\left|0\right\rangle _{2}+\left|1\right\rangle _{0}\left|i\right\rangle _{1}\left|0\right\rangle _{2}\right)$\tabularnewline
$U_{f}(2)$ & \tabularnewline
 & $\frac{1}{\sqrt{2}}\left(\left|0\right\rangle _{0}\left|i\right\rangle _{1}\left|f\right\rangle _{2}+\left|1\right\rangle _{0}\left|i\right\rangle _{1}\left|f\right\rangle _{2}\right)$\tabularnewline
Controlled Swap($0:1\longleftrightarrow2$) & \tabularnewline
 & $\frac{1}{\sqrt{2}}\left(\left|0\right\rangle _{0}\left|i\right\rangle _{1}\left|f\right\rangle _{2}+\left|1\right\rangle _{0}\left|f\right\rangle _{1}\left|i\right\rangle _{2}\right)$\tabularnewline
Hadamard(0) & \tabularnewline
 & $\begin{array}{cc}
\frac{1}{2}\left\{ \left|0\right\rangle _{0}\left(\left|i\right\rangle _{1}\left|f\right\rangle _{2}+\left|f\right\rangle _{1}\left|i\right\rangle _{2}\right)\right.\\
\left.\:+\left|1\right\rangle _{0}\left(\left|i\right\rangle _{1}\left|f\right\rangle _{2}-\left|f\right\rangle _{1}\left|i\right\rangle _{2}\right)\right\} 
\end{array}$\tabularnewline
\end{tabular} 
\par\end{center}

\noindent Now measure the ancillary qubit, register 0. The probability
of measuring this qubit in the state $\left|0\right\rangle $ is
\begin{eqnarray*}
P(0) & = & \frac{1}{4}\left(\left\langle i\right|_{1}\left\langle f\right|_{2}+\left\langle f\right|_{1}\left\langle i\right|_{2}\right)\left(\left|i\right\rangle _{1}\left|f\right\rangle _{2}+\left|f\right\rangle _{1}\left|i\right\rangle _{2}\right),\\
 & = & \frac{1}{2}\left(1+\left\langle i\left|\right.f\right\rangle \left\langle f\left|\right.i\right\rangle \right),
\end{eqnarray*}
 which yields the squared overlap.

The operation of the Hadamard test used to calculate transition amplitudes
is described here. The register 0 is an ancillary qubit and the unitary
operators describing the states and the transition operator $\mathcal{O}$
act on register 1.
\begin{center}
\begin{tabular}{cc}
Operation & State\tabularnewline
\hline 
 & $\left|0\right\rangle _{0}\left|0\right\rangle _{1}$\tabularnewline
Hadamard(0) & \tabularnewline
 & $\frac{1}{\sqrt{2}}\left(\left|0\right\rangle _{0}\left|0\right\rangle _{1}+\left|1\right\rangle _{0}\left|0\right\rangle _{1}\right)$\tabularnewline
Controlled$U_{f}(0,1)$ & \tabularnewline
 & $\frac{1}{\sqrt{2}}\left(\left|0\right\rangle _{0}\left|0\right\rangle _{1}+\left|1\right\rangle _{0}\left|f\right\rangle _{1}\right)$\tabularnewline
PauliX(0) & \tabularnewline
 & $\frac{1}{\sqrt{2}}\left(\left|0\right\rangle _{0}\left|f\right\rangle _{1}+\left|1\right\rangle _{0}\left|0\right\rangle _{1}\right)$\tabularnewline
Controlled$U_{i}(0,1)$ & \tabularnewline
 & $\frac{1}{\sqrt{2}}\left(\left|0\right\rangle _{0}\left|f\right\rangle _{1}+\left|1\right\rangle _{0}\left|i\right\rangle _{1}\right)$\tabularnewline
Controlled$\mathcal{O}(0,1)$ & \tabularnewline
 & $\frac{1}{\sqrt{2}}\left(\left|0\right\rangle _{0}\left|f\right\rangle _{1}+\left|1\right\rangle _{0}\mathcal{O}\left|i\right\rangle _{1}\right)$\tabularnewline
\end{tabular} 
\par\end{center}

Let $\left|\psi\right\rangle $ denote the state $\frac{1}{\sqrt{2}}\left(\left|0\right\rangle _{0}\left|f\right\rangle _{1}+\left|1\right\rangle _{0}\mathcal{O}\left|i\right\rangle _{1}\right)$.
Then
\begin{eqnarray*}
\left\langle \psi\left|\sigma_{x}(0)\right|\psi\right\rangle  & = & \frac{1}{2}\left(\left\langle f\left|\mathcal{O}\right|i\right\rangle +\left\langle i\left|\mathcal{O}\right|f\right\rangle \right),
\end{eqnarray*}
 and 
\begin{eqnarray*}
\left\langle \psi\left|\sigma_{y}(0)\right|\psi\right\rangle  & = & -\frac{i}{2}\left(\left\langle f\left|\mathcal{O}\right|i\right\rangle -\left\langle i\left|\mathcal{O}\right|f\right\rangle \right).
\end{eqnarray*}


\bibliographystyle{h-physrev4}
\bibliography{thebib}

\end{document}